\documentclass[aip,
 amsmath,amssymb,
 reprint,%
]{revtex4-1}

\usepackage{graphicx}
\usepackage{dcolumn}
\usepackage{bm}
\usepackage[utf8]{inputenc}
\usepackage[T1]{fontenc}
\usepackage{mathptmx}
\usepackage{etoolbox}
\usepackage{longtable}

\newcommand\Tstrut{\rule{0pt}{2.6ex}}         
\newcommand\Bstrut{\rule[-1.2ex]{0pt}{0pt}}   

\makeatletter
\def\@email#1#2{%
 \endgroup
 \patchcmd{\titleblock@produce}
  {\frontmatter@RRAPformat}
  {\frontmatter@RRAPformat{\produce@RRAP{*#1\href{mailto:#2}{#2}}}\frontmatter@RRAPformat}
  {}{}
}%
\makeatother
\begin{document}

\preprint{AIP/123-QED}

\title{Spin-polarized alkali-metal trimers revisited}
\author{J. Klime\v{s}}
\author{P. Sold\'{a}n}%
 \email{pavel.soldan@mff.cuni.cz}
\affiliation{Department of Chemical Physics and Optics, Faculty of Mathematics and Physics,
Charles University, Ke Karlovu 3, 121 16 Prague 2, Czech Republic
}%

\date{\today}

\begin{abstract}
Homonuclear spin-polarized alkali-metal trimers in their lowest-lying electronic state are investigated theoretically. Their equilibrium geometries and binding energies are determined with the state-of-the-art quantum chemical methods at three levels of approximation. The equilibrium geometries obtained $R_{\rm eq}({\rm Li}_3) = 3.100$\;\AA, $R_{\rm eq}({\rm Na}_3) = 4.353$\;\AA,  $R_{\rm eq}({\rm K}_3) = 4.996$\;\AA, $R_{\rm eq}({\rm Rb}_3) = 5.391$\;\AA, and  $R_{\rm eq}({\rm Cs}_3) = 5.730$\;\AA\  are compared to the other theoretical results and also with the very recent experimental results obtained through the laser-induced Coulomb explosion. Further theoretical studies are proposed, which could help with better interpretation of the experimental results for the sodium and cesium trimers.   
\end{abstract}

\maketitle

Alkali metals have played pivotal roles in cold and ultracold physics studies. The successful production of alkali-metal dimers in specific states through photoassociation experiments together with the manipulation of scattering properties by exploiting Feshbach resonances opened doors for controlled ultracold chemistry.\cite{Langen-manipulation-2024,Karman-chemistry-2024} It also instigated extensive studies of alkali-metal atom+dimer collision processes, which in turn required good-quality alkali-metal trimer potential energy surfaces. First, the attention focused on the spin-polarized trimers, whose ground electronic states (1$^{4}$A$^{'}_{2}$) are not subject to the Jahn-Teller effect, and their equilibrium geometries exhibit the $D_{3h}$ symmetry (equilateral triangle). In these cases, the corresponding pair-wise additive potentials  were not sufficient because in the alkali-metal trimers the three-body non-additive interactions are very strong. They are shortening  the trimer equilibrium bond lengths by 0.5-1.0 \AA\ with respect to the dimer equilibrium bond lengths and contributing 49-130 \% to the binding energies.\cite{Higgins-Na3-1996,3body-2003,Hetero-2010}  Consequently, potential energy surfaces of the lowest spin-polarized states (1$^{4}$A$^{'}_{2}$) were published for lithium,\cite{Colavecchia-Li3-2003,Li3-2007} sodium,\cite{Higgins-Na3-2000,Na3-2009} potassium,\cite{K3-2005,K3-2008,Hauser-K3-2010} rubidium,\cite{Rb3-2010,Schnabel-Rb3-2021,Bormotova-Rb3-2024} and cesium trimers.\cite{Cs3-2009} It has to be noted that except for Na$_3$ (ref.\;[\onlinecite{Higgins-Na3-1996}]) there were no experimental data for spin-polarized alkali-metal trimers available prior 2000. \\
\indent {{Recently, the importance of three-body interactions in spin-polarized alkali-metal trimers was probed by explaining experimental measurements with theoretical modeling in ultracold spin-polarized NaLi+Na collisions}}.\cite{Park:2023,Karman:2023} Very recently, an experimental study focused on determining the structure of spin-polarized Na$_3$, K$_3$, Rb$_{3}$, and Cs$_3$ on helium nanodroplets through the laser-induced Coulomb explosion.\cite{Experiment-2024} The equilibrium internuclear distances in the 1$^{4}$A$^{'}_{2}$ electronic states determined from the experiment were compared with some of the available theoretical results. Very good agreement (well within the experimental uncertainties) was found for the potassium and rubidium trimers, while a larger discrepancy (0.2-0.3 \AA) for the sodium trimer was attributed to a minor internuclear motion during the laser pulse. In the case of cesium, the observations did not lead to an unambiguous determination of its quantum state. As can be seen in Table \ref{table1}, where the experimental results (second column) and previous theoretical results (third column) are collected,   the spread of the theoretical results is not small, with the exception of the lithium trimer. \\
\indent The goal of our study is to determine the equilibrium inter-nuclear bond lengths of the spin-polarized homonuclear alkali-metal trimers at the highest possible level of theory available to us with the aim of recovering as much of the correlation energy as possible. 
Our approach is based on the minimization of three trimer functions $E^{\mbox{-}{\rm T}}(R)$, $E^{\rm T}(R)$, and $E^{({\rm Q})}(R)$, which describe the interaction energy of homonuclear alkali-metal trimers in equilateral geometries (where all three bond lengths are equal to $R$) with respect to the dissociation limit of three separated atoms and are calculated at three different levels of approximation. For the evaluation of the trimer functions we use the following formulae
\begin{equation}
    E^{\rm T}(R) = E^{\mbox{-}{\rm T}}(R) + \Delta E_{\rm T}(R),
\end{equation}
\begin{equation}
    E^{({\rm Q})}(R) = E^{\rm T}(R) + \Delta E_{({\rm Q})}(R),
\end{equation}
where $E^{\mbox{-}{\rm T}}(R)$ is the interaction energy calculated at the UCCSD-T level on the RHF reference\cite{Deegan:1994} extrapolated to the complete basis 
set limit\cite{Halkier:1998} (CBS) using the aug-cc-pwCVQZ and aug-cc-pwCV5Z basis sets.\cite{Prascher:2011,Hill:2017} $\Delta E_{\rm T}(R)$ is the full-triples\cite{Kallay:2001}  correction term  
\begin{equation}
\Delta E_{\rm T}(R) = E^{\rm T}(R) - E^{\mbox{-}{\rm T}}(R)
\end{equation}
calculated using the aug-cc-pwCVQZ basis set for lithium and 
aug-cc-pwCVTZ for the rest of the elements.\cite{Prascher:2011,Hill:2017}
$\Delta E_{({\rm Q})}(R)$ is the perturbative-quadruples\cite{Kallay:2005,Kallay:2008} correction term 
\begin{equation}
\Delta E_{({\rm Q})}(R) =  E^{({\rm Q})}(R) - E^{\rm T}(R).
\end{equation}
calculated using aug-cc-pwCVTZ for lithium\cite{Prascher:2011}  and aug-cc-pwCVDZ for the other alkali metals.\cite{Hill:2017} 
All the interaction energies calculated are counterpoise corrected for the basis set superposition error.\cite{BSSE} In the case of lithium trimer all nine electrons were correlated. In the case of sodium trimer, 27 electrons were correlated, thus excluding six electrons in the three lowest molecular orbitals. In the cases of potassium, rubidium, and cesium trimers, small-core relativistic pseudopotentials\cite{Lim:2005} were used to describe inner-core electrons, and all 27 electrons remaining as the large valence were correlated.
The UCCSD-T energies were calculated using the MOLPRO program suite\cite{MOLPRO-WIREs, MOLPRO-JCP} and the correction terms using the MRCC program suite.\cite{MRCC-JCP} (Recently, a similar approach was used {{to study reactions of alkaline-earth-metal  diatomic molecules \cite{Tomza-reactions:2023}}} and to determine the interaction potentials of diatomic molecules consisting of alkali-metal and alkaline-earth-metal atoms.\cite{Tomza-Dimers:2024}) For each species, the CBS interaction energies $E^{\mbox{-}{\rm T}}(R)$ and trimer functions $E^{\rm T}(R)$, and $E^{({\rm Q})}(R)$ were evaluated on a grid of several points near the suspected minimum. From these, the actual minimum bond lengths $R_{\rm eq}^{\mbox{-}{\rm T}}$, $R_{\rm eq}^{\rm T}$, and $R_{\rm eq}^{({\rm Q})}$ were determined. Then single-point calculations were performed at the minimum geometries ($R_{\rm eq}^{({\rm Q})}$) for each species to determine individual contributions to the interaction energies.

\begin{table*}[t]
\caption{\label{table1} $D_{3h}$ equilibrium bond lengths (\AA)  for the homonuclear spin-polarized alkali-metal trimers in the 1$^{4}$A$^{'}_{2}$ electronic state.}
\begin{ruledtabular}
\begin{tabular}{llcccccc} & Experiment$^a$ & Other theory & $R_{\rm eq}^{\mbox{-}{\rm T}}$
 & $R_{\rm eq}^{\rm T}$ & $R_{\rm eq}^{({\rm Q})}$ & T1 \Bstrut \\
\hline \Tstrut
 Li$_{3}$ & -               & 3.103,$^{b}\;$ 3.09,$^{c}\;$ 3.1016,$^{d}\;$ 3.102$^{e}\;$ & 3.098 & 3.100 & $\mathbf{3.100}$  & 0.025 \Bstrut\\
Na$_{3}$ & 4.65 $\pm$ 0.15 & 4.428,$^{b}\;$ 4.33,$^{c}\;$  4.406,$^{f}\;$ 4.34$^{g}$ & 4.369 & 4.352 & $\mathbf{4.353}$  & 0.009 \Bstrut \\ 
K$_{3}$  & 5.03 $\pm$ 0.18 & 5.084,$^{b}\;$ 4.95,$^{c}\;$  5.09,$^{h}\;$ 5.05$^{i}$ & 5.007 & 5.000 & $\mathbf{4.996}$  & 0.018 \Bstrut\\ %
Rb$_{3}$ & 5.45 $\pm$ 0.22 & 5.596,$^{b}\;$ 5.35,$^{c}\;$ 5.50,$^{i}\;$ 5.45,$^{j}\;$ 5.311$^{k,l}$ & 5.416 & 5.395& $\mathbf{5.391}$ & 0.020 \Bstrut \\
Cs$_{3}$ & -               & 5.992,$^{b}\;$ 5.56,$^{c}\;$ 5.92,$^{m}\;$ 5.67,$^{m}\;$ & 5.708& 5.734 & $\mathbf{5.730}$ & 0.041 \\
\end{tabular}
\end{ruledtabular}
$^{a}\;${Ref.~\onlinecite{Experiment-2024},}
$^{b}\;${Ref.~\onlinecite{3body-2003},}
$^{c}\;${Ref.~\onlinecite{Hetero-2010},}
$^{d}\;${Ref.~\onlinecite{Colavecchia-Li3-2003},}
$^{e}\;${Ref.~\onlinecite{Li3-2007},}
$^{f}\;${Ref.~\onlinecite{Higgins-Na3-2000},}
$^{g}\;${Ref.~\onlinecite{Na3-2009},}
$^{h}\;${Ref.~\onlinecite{K3-2005},}
$^{i}\;${Ref.~\onlinecite{Hauser-K3-2010},}
$^{j}\;${Ref.~\onlinecite{Rb3-2010},}
$^{k}\;${Ref.~\onlinecite{Schnabel-Rb3-2021},}
$^{l}\;${Ref.~\onlinecite{Bormotova-Rb3-2024},}
$^{m}\;${Ref.~\onlinecite{Cs3-2009}.}
\end{table*}

Our resulting equilibrium bond lengths are collected in the fourth, fifth, and sixth columns of Table \ref{table1}, while the contributions to the corresponding interaction energies are collected in the second, third, and fifth columns of Table \ref{table2}. Generally, the effect of the perturbative-quadruples corrections on the equilibrium bond lengths is much smaller than that of the full-triples corrections. 
{{However, it has to be emphasized that we were able to use only small basis sets for the evaluation of the perturbative-quadruples corrections for heavier alkali metals.}}

Our equilibrium bond lengths for K$_3$ and Rb$_3$ agree well with the other theoretical results and thus are also very close to the values determined experimentally. In the case of Na$_3$, the newly calculated bond length 4.344 \AA\ is still significantly shorter than the lower bound 4.50 \AA\ determined by the error analysis of the experiment, although it agrees very well with the results of Refs.\;[\onlinecite{Hetero-2010}] and [\onlinecite{Na3-2009}], where completely different \textit{ab initio} methods were used. A possible explanation for this experiment-theory discrepancy, as suggested in Ref.\;[\onlinecite{Experiment-2024}], is a minor internuclear motion during the laser pulse due to the lightness of sodium. In the future, this could be verified by calculations of the mean values of $R$ in relevant excited vibrational states employing one of the available potential energy surfaces.\cite{Higgins-Na3-2000,Na3-2009} 

For Cs$_3$, the comparison with the experiment is even more complicated. The experimental measurements were discussed; however, no equilibrium bond lengths were reported in the article.\cite{Experiment-2024} The corresponding kinetic energy release exhibits two peaks, each dominating at different nozzle temperatures, which under the assumption of $D_{3h}$ geometry would correspond to 6.10 \AA\ or 5.60 \AA. 
Our calculated value is not close to any of them. If one eliminates the internuclear motion for the heaviest of the studied systems, it appears that other electronic states are involved. Thus, to find viable candidates, a detailed \textit{ab initio}  multi-reference study of low-lying electronic states of the cesium trimer including doublet and quartet spin multiplicities will be required (obviously lowering the $D_{3h}$ symmetry restriction to $C_{2v}$). In addition to further calculations, it may also require another look at the experimental data to determine whether the $D_{3h}$ symmetry assumption has been adequate in this case. 

The above conclusions are also partly supported by the T1 characteristics\cite{T1:1989} collected for all species in the fifth column of Table \ref{table1}. In \textit{ab initio} calculations, the T1 diagnostic helps to assess the adequacy of a single-reference approach. Values higher than 0.02 usually indicate that a multi-reference electron correlation treatment is needed. The Cs$_3$ T1 value is by far the worst of all species considered, while Na$_3$ appears to be the best suited for single-reference methods.

\begin{figure}
   \centering
   \includegraphics[width=6cm,angle=270]{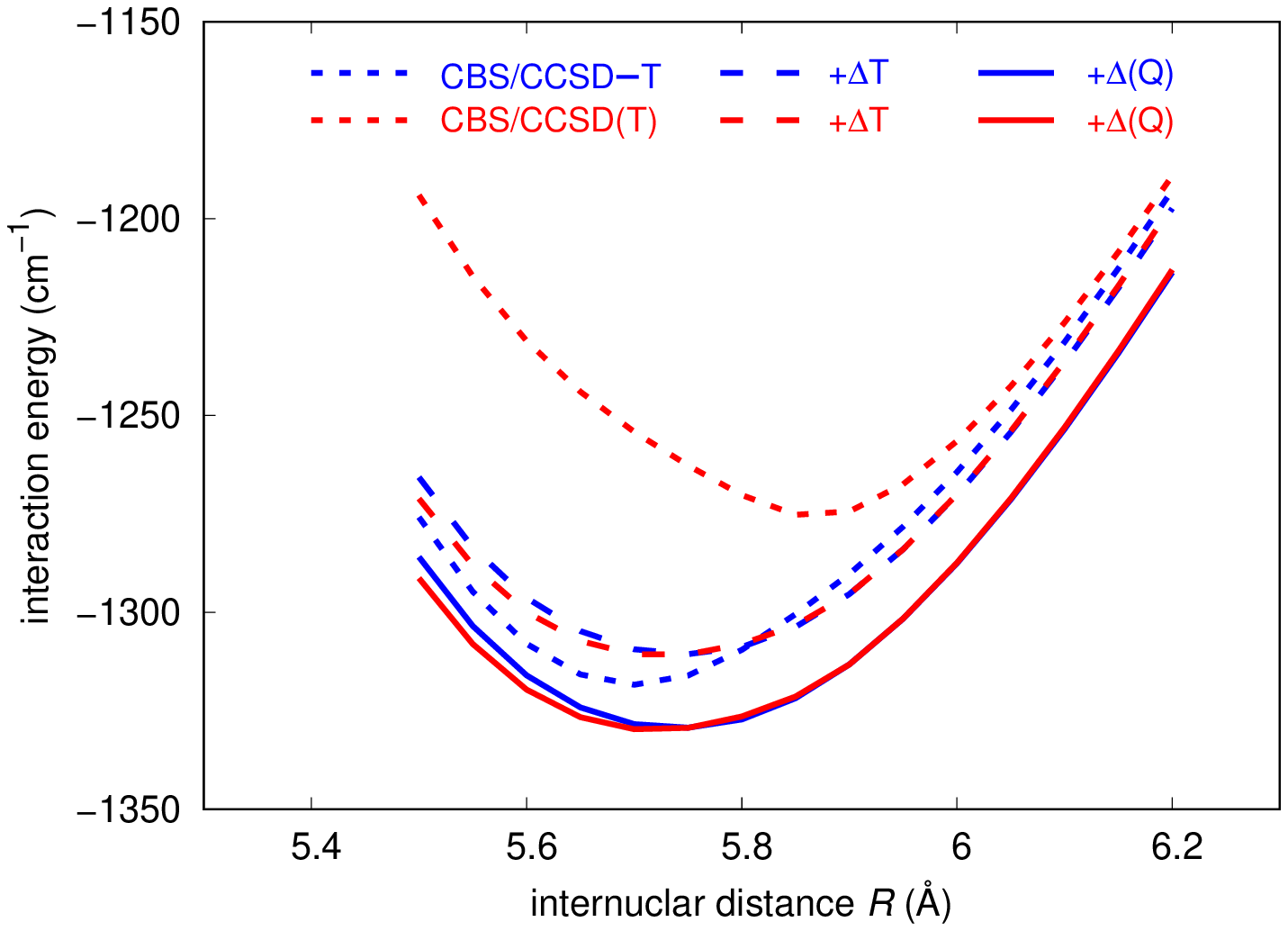}
      \caption{Comparison of trimer functions for the cesium trimer based on the CBS CCSD(T) (red) and CCSD-T (blue) interaction energies (dotted lines) including the full-triples correction (dashed lines) and the perturbative-quadruples correction (full lines). 
              }
         \label{cstrim}
\end{figure}

The cesium trimer is also the reason we used the CCSD-T method rather than the very popular CCSD(T) method as a base for our CBS extrapolation. In Fig. \ref{cstrim}, there is a comparison of the CBS interaction energies extrapolated from the CCSD-T and CCSD(T) values together with the corresponding trimer functions. The CBS CCSD(T) minimum is at a much longer distance (5.816 \AA) than the CCSD-T minimum. The full-triples corrections move both these minima to nearly the same distance, which is actually much closer to the CCSD-T minimum. Although in other alkali metals we saw much smaller differences between the positions of their CBS CCSD-T and CCSD(T) minima (the largest difference is 0.023 \AA\ for K$_3$), for consistency, we decided to use the CCSD-T method as a base for our CBS extrapolation.

\begin{table*}[t]
\caption{\label{table2} Individual contributions to the equilibrium interaction energies$^{a}$ (cm$^{-1}$) for the homonuclear spin-polarized alkali-metal trimers in the 1$^{4}$A$^{'}_{2}$ electronic state.}
\begin{ruledtabular}
\begin{tabular}{lrrrrrrr}
 & $E^{\mbox{-}{\rm T}}$ & $\Delta E_{\rm T}$ & $E^{\rm T}$
 & $\Delta E_{({\rm Q})}$ & $E^{({\rm Q})}$ & \multicolumn{2}{c}{Other theory} \Bstrut \\
\hline \Bstrut
\Tstrut  Li$_{3}$ &  $-4010$& $-55$ & $-4065$ & $-2$ & $\mathbf{-4067}$ & -4022,$^{b}\;$ & -4315,$^{c}\;$ \Bstrut \\
Na$_{3}$ &  $-854$ & $-29$ & $-883$  & {$1$} & {$\mathbf{-882}$}  &  -837,$^{b}\;$ & -839,$^{c}\;$ \Bstrut \\ 
K$_{3}$  &  $-1253$& $-26$ & $-1279$ & $-13$ & $\mathbf{-1292}$ & -1274,$^{b}\;$ & -1313,$^{c}\;$ \Bstrut \\ %
Rb$_{3}$ &  $-1127$& $-18$ & $-1145$ & {$-15$} & {$\mathbf{-1160}$} &  -995,$^{b}\;$ & -1103,$^{c}\;$ \Bstrut \\
Cs$_{3}$ &  {$-1318$}& $7$   & $-1311$ & {$-19$} & {$\mathbf{-1330}$} & -1139,$^{b}\;$ & -1456,$^{c}\;$ \\
\end{tabular}
\end{ruledtabular}
$^{a}\;$Calculated at $R_{\rm eq}^{({\rm Q})}$ with respect to the three-separated-atoms dissociation limit.
$^{b}\;${Ref.~\onlinecite{3body-2003},}
$^{c}\;${Ref.~\onlinecite{Hetero-2010}.}
\end{table*}

\begin{table*}[ht]
{{
\caption{\label{table3} Additive two-body and non-additive three-body contributions to UCCSD-T interaction energies$^{a}$ (cm$^{-1}$) and their respective corrections for the homonuclear spin-polarized alkali-metal trimers in the 1$^{4}$A$^{'}_{2}$ electronic state.}
\begin{ruledtabular}
\begin{tabular}{llccccc}
Method& Component & Li$_3$  & Na$_3$ & K$_3$ & Rb$_3$ & Cs$_3$ \Bstrut \\
\hline 
\Tstrut CCSD-T& $E^{\mbox{-}{\rm T}}$ & -4010 & -854 & -1253 & -1127 & -1318 \Bstrut \\
    & additive two-body contribution & 1242 & -123 & -390 & -459 & -552 \Bstrut \\
    & non-additive three body contribution & -5252 & -731 & -863 & -668 & -766 \Bstrut \\
Other theory    & non-additive three body contribution$^{b}$ & -5260 & -663 & -831 & -513 & -562 \Bstrut \\
    & non-additive three body contribution$^{c}$ & -5559 & -768 & -958 & -723 & -990 \Bstrut \\
$\Delta$CCSDT& $\Delta E_{\rm T}$ &-55 & -29 & -26 & -18 & 7 \Bstrut \\
    &additive two-body contribution&-12 & -9 & -3 & -3 & 6 \Bstrut \\
    &non-additive three body contribution &-43 & -20 & -23 & -15 & 1 \Bstrut \\
$\Delta$CCSDT(Q)& $\Delta E_{({\rm Q})}$ &-2 & 1 & -13 & -15 & -19 \Bstrut \\
    &additive two-body contribution&0 & 1 & -9 & -12 & -15 \Bstrut \\
    &non-additive three body contribution & -2 & 0 & -4 & -3 & -4 \Bstrut\\
\end{tabular}
\end{ruledtabular}
$^{a}\;$Calculated at $R_{\rm eq}^{({\rm Q})}$ with respect to the three-separated-atoms dissociation limit.
$^{b}\;${Ref.~\onlinecite{3body-2003},}
$^{c}\;${Ref.~\onlinecite{Hetero-2010}.}
}}
\end{table*}

{{Contribution of non-additive three-body forces to the corrections of the interaction energies is another aspect worth investigating. Additive two-body and non-additive three-body contributions to the UCCSD-T interaction energies and their corrections (calculated at $R_{\rm eq}^{({\rm Q})}$) are presented in Table \ref{table3}. As expected, the non-additive three-body contributions to the UCCSD-T interaction energies follow the same pattern as in previous systematic studies.\cite{3body-2003, Hetero-2010}  Because these contributions make majority of the binding energies, the interaction energies follow the same pattern. Except for the cesium trimer, the non-additive three-body contribution is the major one in the full-triples correction terms. Interestingly enough, with the exception of the lithium trimer the additive two-body contribution dominates the perturbative-quadruple correction terms.}}

In summary, we determined the equilibrium bond lengths for the homonuclear spin-polarized alkali-metal trimers in the lowest quartet state 1$^{4}$A$^{'}_{2}$ at three levels of approximation with the \textit{ab initio} calculations performed at the highest possible level available to us. Further theoretical studies were proposed, which could help with better interpretation of the experimental results for the sodium and cesium trimers.

\section*{AUTHOR DECLARATIONS}

 The authors have no conflicts to disclose.
\section*{DATA AVAILABILITY}
 The data that support the findings of this study are available on {Zenodo, DOI: 10.5281/zenodo.15274246}. 

\begin{acknowledgements}
Computational resources were provided by the e-INFRA CZ project (ID:90254),
supported by the Ministry of Education, Youth and Sports of the Czech Republic, and
by the ELIXIR-CZ project (ID:90255), part of the international ELIXIR infrastructure.
JK thanks to the support of the Advanced Multiscale Materials for Key Enabling Technologies project, 
supported by the Ministry of Education, Youth, and Sports of the Czech Republic, 
project No. CZ.02.01.01/00/22\_008/0004558, co-funded by the European Union.
\end{acknowledgements}


\providecommand{\noopsort}[1]{}\providecommand{\singleletter}[1]{#1}%

\end{document}